\documentclass[10pt,a4paper,twoside]{article}

\usepackage{indentfirst} 
\usepackage{graphicx} 
\usepackage{amsgen,amsfonts,amssymb,amsbsy} 

\newenvironment{resum}{\begin{quote}\small}{\end{quote}}

\newcommand{\bfsf}[1]{\textsf{\textbf{#1}}}

\begin{document}

\thispagestyle{plain} 

\begin{center}


{\LARGE\bfsf{Quantum Singularities}}

\bigskip


\textbf{D.A. Konkowski}$^1$, \\  \textbf{T.M. Helliwell}$^2$, and 
\textbf{C. Wieland}$^2$


$^1$\textsl{U.S. Naval Academy, U.S.A. and Queen Mary, University of 
London, U.K.} \\
$^2$\textsl{Harvey Mudd College, U.S.A.}

\end{center}

\medskip


\begin{resum}
The definitions of classical and quantum singularities in general 
relativity are reviewed. The occurence of quantum mechanical
singularities in certain spherically symmetric and cylindrically symmetric
(including infinite line mass) 
spacetimes is considered. A strong repulsive ``potential'' near the
classical singularity is shown to turn a classically singular spacetime 
into a quantum mechanically nonsingular spacetime.
\end{resum}

\bigskip


\section{Introduction}

In classical general relativity singularities are not part of the 
spacetime; they are boundary points indicated by incomplete geodesics in
an otherwise maximal spacetime. Thus, at least for timelike and null 
geodesics, this incompleteness can be considered as the abrupt ending of
classical particle paths. What happens if, instead of classical
particles, quantum mechanical wave packets are used? This is
the question G. Horowitz and D. Marolf \cite{hm} set out to answer. We 
will review their definition here and use simple spherical and 
cylindrical spacetimes to illustrate the effects drawing a conclusion, 
in the end, about which classical singularities turn quantum 
mechanically nonsingular.

\section{Classical Singularities}

A spacetime is defined as a connected, $C^{\infty}$, paracompact, Hausdorff
manifold $M$ with Lorentzian metric $g_{\mu\nu}$ \cite{he}. A classical
singularity in a maximal spacetime is indicated by incomplete geodesics 
and/or incomplete curves of bounded acceleration \cite{es}.

Singularities in maximal spacetimes can be classified \cite{es} 
into three basic types: quasiregular, nonscalar curvature, and scalar 
curvature. The mildest is quasiregular and the strongest is scalar 
curvature. At a scalar curvature singularity, physical quantities such as 
energy density and tidal forces diverge in the frame of all observers who 
approach the singularity. For example, the center of a Schwarzschild black 
hole or the beginning of a Big Bang cosmology \cite{es} \cite{he}. At a 
nonscalar curvature singularity, there 
exist curves through each point arbitrarily close to the singularity such 
that observers moving on these curves experience perfectly regular tidal 
forces. Whimper cosmologies are a good example \cite{ek}.  For a 
quasiregular singularity, no obserevers see physical 
quantities diverge, even though their worldlines end at the singularity in 
a finite proper time. The canonical examples are conical singularities 
(e.g., 2D cones \cite{es} and idealized cosmic strings \cite{fv}).

\section{Quantum Singularities}

To decide whether a spacetime is quantum mechanically singular we will 
use the
criterion proposed by Horowitz and Marolf \cite{hm}. They call a spacetime 
quantum mechanically {\it non}singular if the evolution of a test wave 
packet 
in the spacetime is uniquely determined by the initial wave packet, without 
having to put arbitrary boundary conditions at the classical singularity. 
Their construction is restricted to static spacetimes.

According to Horowitz and Marolf, a static spacetime is quantum 
mechanically 
singular if the spatial portion of the Klein-Gordon wave operator is not 
essentially self-adjoint \cite{rs}. A relativistic scalar quantum particle 
with mass $M$ can be described by the positive frequency solution to the 
Klein-Gordon equation $\frac{\partial^2\psi}{\partial{t^2}}=-A\psi$
in a static spacetime where the spatial operator $A$ is defined to be
$A=-VD^{i}(VD_{i})+V^{2}M^{2}$ with $V=-\xi_{\nu}\xi^{\nu}$. Here 
$\xi^{\nu}$ is the timelike Killing field and $D_{i}$ is the spatial 
covariant derivative on the static slice $\Sigma$. The appropriate Hilbert 
space is $L^{2}(\Sigma)$, the space of square integrable functions on 
$\Sigma$. 

If we initially define the domain of $A$ to be 
$C_{0}^{\infty}(\Sigma)$, $A$ is real, positive, symmetric operator and 
self-adjoint extensions always exist
\cite{rs}. If there is only a single, unique extension $A_{E}$, the $A$ is 
essentially self-adjoint. In this case, the Klein-Gordon equation for a free 
scalar particle takes the form \cite{hm}:  $i\frac{d\psi}{dt}=A_E^{1/2}\psi$
with $\psi(t)=exp(-it(A_E)^{1/2})\psi(0)$.
These equations are ambiguous if $A$ is not essentially self adjoint. This 
fact led Horowitz and Marolf to define 
quantum mechanically singular spacetimes as those in which $A$ is not 
essentially self-adjoint. Examples are considered by Horowitz and Marolf
\cite{hm}.

A simple test for essential self-adjointness of the operator (i.e.,
quantum singularity of the spacetime) may be used \cite{es} \cite{rs}.
Essentially one takes the solutions to a test equation

\begin{equation}
(\bigtriangledown ^2 \pm i)\Phi= 0
\label{eq.1}
\end{equation}
and looks to see whether there is more than one $L^2$ solution for each $i$ and each choice of separation constant, near the singularity. If there is
more than one $L^2$ solution the spacetime is quantum mechanically
singular.

\section{Spherical Spacetimes}

A class of static, spherical spacetimes with timelike singularities were 
considered  first in the Horowitz and Marolf paper \cite{hm}. Consider 
the metric 

$$ds^2 = -dt^2+dr^2+r^{2p}(d \theta ^2+\sin ^2 \theta d \phi^2).$$
This spacetime is geodesically incomplete and thus classically singular 
unless 
$p=1$. What about quantum mechanically singular? Take the test equation 
Eq.(1),
separate variables,
$\Psi \sim f(r)Y(\theta ,\phi)$, and consider the radial equation, 
$$f'' + \frac{2p}{r} f' + [ \pm i - \frac{c}{r^{2p}}] f = 0$$
where c is a constant. Next rewrite the radial equation in 
Schrodinger form by letting $f=r^{-p}F$ and obtain
$$F'' + [ \pm i - \frac{p(p-1)}{r^2} - \frac{c}{r^{2p}}] F=0.$$
Near $r=0$, if $0<p<1$ the ``potential'' is attractive, while if 
$p \ge 1$ the ``potential'' is repulsive.
Near $r=0$, 
one solution of the original equation goes like a constant (and is thus 
$L^2$ using the appropriate measure) and the other goes like
$r^{1-2p}$ (and is thus $L^2$ if $p<\frac{3}{2}$). We, therefore, see that 
these spherical spacetimes are quantum mechanically singular, if
$p<\frac{3}{2}$ 
(unless $p=1$), and quantum mechanically singular if $p>\frac{3}{2}$ (or
$p=1$). The spacetimes are quantum mechanically nonsingular if the spacetime 
metric induces a very {\bf repulsive} potential.

\section{Cylindrical Spacetimes}

Next consider a class of static, cylindrical metrics with timelike 
singularities,
$$ds^2 = -dt^2 +dr^2 + r^{2a}d \phi ^2 + r^{2b}dz^2,$$
where $r$ is a radial coordinate and $\phi$ is an angular coordinate with 
the 
usual ranges. These
metrics are geodesically incomplete and thus classically singular unless 
$a=1$ and $b=0$ (flat spacetime in cylindrical polar coordinates). What about 
quantum mechanically singular? Take the test equation Eq.(1), separate 
variables so 
$\Psi\sim f(r)\exp (im \phi) \exp (ikz)$ , and consider the radial equation,
$$f'' = \frac{a+b}{r} f' + [ \pm i - \frac{m^2}{r^{2a}} - 
\frac{k^2}{r^{2b}}]f =0.$$
We can rewrite the radial equation in Schodinger form by letting
$f=r^{(a+b)/2}F$ and obtain
$$F'' + [ \pm i - \frac{(\frac{a+b}{2})[(\frac{a+b}{2}) -1]}{r^2}
-\frac{m^2}{r^{2a}} - \frac{k^2}{r^{2b}}]F=0.$$
Assume $m=0$, $k=0$, for simplicity. Near $r=0$, if $a+b<2$, the ``potential''
is attractive, while if $a+b \ge 2$, the ``potential'' is repulsive.

Near $r=0$, one solution of the original equation goes like a constant 
(and is thus $L^2$ 
using
the appropriate measure) and the other goes like $r^{1-(a+b)}$ (and is thus 
$L^2$ if $a+b<3$). We, therefore, see that these cylindrical spacetimes are 
quantum mechanically singular if $a+b<3$ (except for Minkowski spacetime) 
and quantum mechanically nonsingular if $a+b \ge 3$ (or Minkowski 
spacetime). 
These cylindrical spacetimes are thus quantum mechanically nonsingular if 
the spacetime metric induces a very {\bf repulsive} potential.

\section{``Infinite Line Mass'' Spacetimes}

Finally, consider another class of cylindrical spacetimes. For certain 
parameter values one can interpret the Levi-Civita metric,
$$ds^2 = r^{4\sigma }dt^2 - r^{8\sigma ^2 -4\sigma }(dr^2 +dz^2) -
\frac{r^{2-4\sigma }}{c^2}d\theta ^2.$$
as an ``infinite line mass'' spacetime. In fact, after some controversy in the
literature (see, e.g. \cite{bon}, \cite{hrs}), the following interpretations 
have become somewhat accepted: $\sigma =0,1/2$  locally flat; 
$\sigma =0,c=1$ Minkowski spacetime;  $\sigma =0,c \ne 1$  cosmic string 
spacetime; $0<\sigma <1/2$  infinite line mass spacetimes;
$\sigma = 1/2$ Minkowski spacetime in accelerated 
coordinates (planar source). 

The Levi-Civita metric is static, cylindrically symmetric and 
classically singular at $r=0$ unless (a) $\sigma =0, c=1$ or 
(b) $\sigma =1/2$.
What about quantum mechanically singular? One can again use the test 
equation, separate variables and obtain a radial equation which can be 
written in 
Schrodinger form. (For brevity we will just give the results here; details 
can be found in Konkowski, Helliwell and Wieland \cite{khw}). 
We find that both
linearly independent solutions to the second order ordinary differential 
radial equation are $L^2$ at $r=0$ except: (a) $\sigma =0, c=1$, any $m$, 
Minkowski spacetime, (b) $\sigma =0, |m|C \ge 1$, 
cosmic string spacetime (thus cosmic string is ``generically'' singular 
for wave packets which are a combination of arbitrary modes 
\cite{kh} \cite{khw}), and $\sigma =1/2$, any m, Minkowski spacetime in accelerated 
coordinates. Again, a strong repulsive potential for certain modes shields 
the singularity 
(in the cosmic string case) from the conical singularity on the axis.

All Levi-Civita spacetimes are thus quantum mechanically singular 
except 
Minkowski ($\sigma =0, c=1$) and Minkowski in accelerated 
coordinates 
($\sigma =1/2$). 
One needs to interpret the `quantum singularity' of the physically 
reasonable 
infinite line mass spacetimes as the need for one to put 
``boundary conditions''
at the line mass itself or `round-off' this $\delta$-function singularity 
and 
then put boundary conditions at the matter surface (a simlar argument was 
used for the cosmic string 
case, see e.g., Kay and Studer \cite{ks}).

\section{Conclusions}

It is thus clear that if the repulsive barrier near the classical 
singularity is {\it sufficiently strong}, the probability of a quantum 
mechanical particle penetrating to the origin is {\it sufficiently small},
that the quantum mechanical particle doesn't feel the singularity in
some sense.

One can ask whether the ``strength'' of a classical singularity has any 
effect on the existence of a quantum mechanical singularity. The answer 
appears to be no: spacetimes with classical quasiregular singularities 
are as likely to be quantum mechanically singular as are classical 
scalar curvature singularities \cite{kh}.

One can also ask whether the type of probing particle (scalar, null vector, spinor) has any effect on whether a singularity is or is not quantum mechanically singular. The answer appears to be generically no, although the wave packet modes producing the the quantum singularity may differ depending on particle type \cite{hka}.


\end{document}